\documentclass[prl, twocolumn,doublespaced,preprintnumbers,amsmath,amssymb]{revtex4}
\usepackage[english]{babel}
\usepackage{amsmath,amsfonts,amssymb}
\usepackage[pdftex]{graphicx}
\usepackage{dcolumn}
\usepackage{bm}

\begin{document}
\title{Dense Electron-Positron Plasmas and Ultra-Intense Bursts of Gamma-Rays from Laser-Irradiated Solids}
\author{C.P. Ridgers$^{1,2}$, C.S. Brady$^3$, R. Duclous$^4$, J.G. Kirk$^5$, K. Bennett$^3$, T.D. Arber$^3$, A.P.L. Robinson$^2$, A.R. Bell$^{1,2}$}

\address{{\small{\small{$^1$Clarendon Laboratory, University of Oxford, Parks Road, Oxford, OX1 3PU, UK \\
$^2$Central Laser Facility, STFC Rutherford-Appleton Laboratory, Chilton, Didcot, Oxfordshire, OX11 0QX, UK \\
$^3$Centre for Fusion, Space and Astrophysics, University of Warwick, Coventry, CV4 7AL, UK \\
$^4$Commissariat \`{a} l'Energie Atomique, DAM DIF, F-91297 Arpajon, France  \\
$^5$Max-Planck-Institut f\"{u}r Kernphysik, Postfach 10 39 80, 69029 Heidelberg, Germany}}}}

\begin{abstract}
In simulations of a 10PW laser striking a solid we demonstrate the possibility of producing a pure electron-positron plasma by the same processes as those thought to operate in high-energy astrophysical environments. A maximum positron density of $10^{26}$m$^{-3}$ can be achieved, seven orders of magnitude greater than achieved in previous experiments.  Additionally, 35\% of the laser energy is converted to a burst of gamma-rays of intensity $10^{22}$Wcm$^{-2}$, potentially the most intense gamma-ray source available in the laboratory.  This absorption results in a strong feedback between both pair and $\gamma$-ray production and classical plasma physics in the new `QED-plasma' regime.
\end{abstract}

\maketitle


Electron-positron ($e^-e^+$) plasmas are a prominent feature of the winds from pulsars and black holes \cite{Goldreich_69,Blandford_77}.  They result from the presence of electromagnetic fields strong enough to cause non-linear quantum electrodynamics (QED) reactions \cite{Ritus_85} in these environments leading to a cascade of $e^-e^+$ pair production \cite{Timhokin_10}.  These fields can be much lower than the Schwinger field for vacuum breakdown \cite{Schwinger_51} if they interact with highly relativistic electrons ($\gamma>>1$) \cite{Ritus_85}.  Non-linear QED has been probed experimentally with lasers in two complementary ways: (1) with a particle accelerator accelerating electrons to the necessary $\gamma$ and a laser supplying the fields \cite{Burke_97,Sokolov_10,Hu_10}; or (2) with a laser accelerating the electrons and gold-nuclei supplying the fields \cite{Chen_10,Liang_98,Shen_01}.  An alternative configuration, using next-generation high-intensity lasers to provide both the acceleration and the fields \cite{Bell_08}, has the potential to generate dense $e^-e^+$ plasmas.  Analytical calculations and simulations exploring this configuration have shown that an overdense $e^-e^+$ plasma can be generated from a single electron by counter-propagating 100PW lasers \cite{Bell_08,Fedotov_10,Bulanov_10,Nerush_11}.  Here we will show that such a plasma can be generated with an order of magnitude less laser power by firing the laser at a solid target, putting such experiments in reach of next-generation 10PW lasers \cite{Mourou_07}.  
 
The dominant non-linear QED effects in 10PW laser-plasma interactions are: synchrotron gamma-ray photon ($\gamma_h$) emission from electrons in the laser's electromagnetic fields; and pair-production by the multiphoton Breit-Wheeler process, $\gamma_h+n\gamma_l\rightarrow e^- + e^+$, where $\gamma_l$ is a laser photon \cite{Kirk_09,Ritus_85,Erber_66}.  Each reaction is a strongly multiphoton process, the former process being non-linear Compton scattering, $e^-+m\gamma_l\rightarrow e^-+\gamma_h$ \cite{Mackenroth_11,DiPiazza_10}, in the limit $m\rightarrow\infty$.  Therefore, these reactions only become important at the ultra-high intensities reached in 10PW laser-plasma interactions.  The importance of synchrotron emission is determined by the parameter $\eta$.  This depends on the ratio of the electric and magnetic fields in the plasma to the Schwinger field \cite{Schwinger_51} ($E_s=1.3\times10^{18}$Vm$^{-1}$).  For ultra-relativistic particles $\eta=(\gamma/E_s)|\mathbf{E}_{\perp}+\underline{\beta}\times{}c\mathbf{B}|$ \cite{Kirk_09,Erber_66}.   $\gamma$ is the Lorentz factor of the emitting electron or positron, $\underline{\beta}$ is the corresponding velocity normalised to $c$ and $\mathbf{E}_{\perp}$ is the electric field perpendicular to its motion.  As $\eta$ approaches unity each emitted photon takes a large fraction ($\approx0.44$) of the emitting electron's energy, and the mean free path of these photons to pair production is of the order of the laser wavelength so that many pairs are produced \cite{Bell_08}.  For a 10PW laser operating at an intensity of $10^{23}$Wcm$^{-2}$ $|\mathbf{E}|\approx10^{15}\mbox{Vm}^{-1}$.  On interacting with a plasma such a laser accelerates electrons to a $\gamma$ of the order of several hundreds and so $\eta$ approaches one.  However, the geometry of the interaction is crucial; for a single intense laser beam striking a single electron, the electron is rapidly accelerated to $\approx c$ in the direction of propagation of the laser pulse.  In this case $\mathbf{E}_{\perp}$ is almost exactly cancelled by $\mathbf{v}\times\mathbf{B}$, $\eta$ is reduced and pair production is dramatically curtailed.  By contrast, in an overdense plasma the wave becomes evanescent and the terms do not cancel.  Therefore, laser-solid interactions offer an attractive route to generating electron-positron plasmas. 

In this Letter we will present the first simulations of 10PW laser-solid interactions to include the relevant QED processes.  We show that such interactions are the most effective way to produce $e^-e^+$ pairs with next-generation lasers and that the laser is absorbed into an ultra-intense burst of $\gamma$-rays with high efficiency (35\%).  In order to understand these interactions it is crucial to resolve the complex feedback between QED and collective plasma physics effects.  Therefore, in contrast to the schemes described in (1) \& (2) above, we describe a new regime in which QED processes and plasma physics are inseparable, which we term a `QED-plasma'.  

In order to simulate QED-plasmas we have included synchrotron emission of high energy $\gamma$-ray photons and Breit-Wheeler pair production in the particle-in-cell (PIC) code {\small{\small{EPOCH}}} \cite{Brady_11}. As $\eta$ approaches unity the high energy of the emitted photons means that radiation must be considered discontinuously.  The electrons and positrons obey the Lorentz force equation, following the classical worldlines as computed by the PIC code, until a discrete photon is emitted \cite{Shen_72}.  The recoil in such an event provides a discontinuous radiation reaction force \cite{DiPiazza_10}.  As discussed below, the discontinuous radiation model consists of random sampling of the synchrotron spectrum and so tends to the continuous-loss model \cite{Dirac_38,Landau_87,Sokolov_09,Harvey_11,Kirk_09} as $\hbar\omega_h<<\gamma{}m_ec^2$ ($\hbar\omega_h$ is the energy of the emitted photon), i.e. as the sampling frequency $\rightarrow\infty$.  It has recently been shown that in 10PW laser-plasma interactions the discontinuous model yields an order of magnitude more $e^-e^+$ pairs \cite{Duclous_11}.  This is due to some electrons reaching higher energies and emitting a higher energy photon than the same electron experiencing a continuous radiation drag force, the so-called `straggling' effect \cite{Shen_72}.  

\begin{figure}
\centering
\includegraphics[scale=0.4]{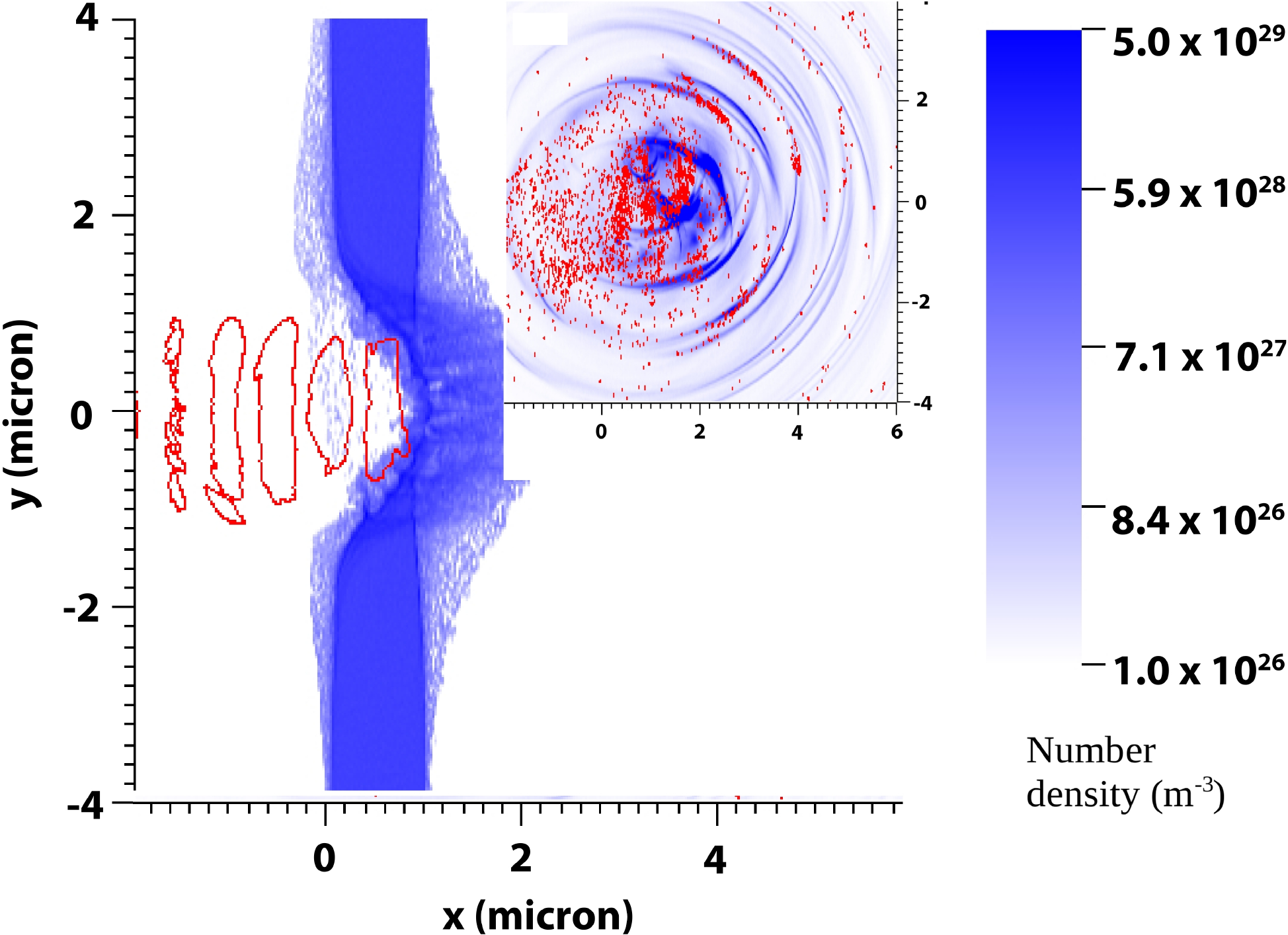}
\caption{\label{solid_target_2D} Pair-production by a laser of intensity $4\times10^{23}$Wcm$^{-2}$ striking an aluminium target (snapshots at the end of the 30fs laser pulse).  The laser (red contours) bores a hole into the solid target (blue).  Gamma-rays (blue) and positrons (red dots) are generated in this interaction (inset -- on the same scale).}
\end{figure}

The QED processes are simulated using a Monte-Carlo algorithm \cite{Duclous_11}.  The time at which emission events occur is computed as follows.  Each particle is assigned an optical depth at which it emits ($\tau$) according to $P=1-e^{-\tau}$, where $P\in[0,1]$ is chosen at random to capture the quantum fluctuations in the emission processes and so the straggling.  The rates of photon and pair production, $d\tau_{\gamma}/dt=(\sqrt{3}\alpha_fc\eta)/(\lambda_C\gamma)\int^{\eta/2}_0d\chi{}F(\eta,\chi)/\chi$ and $d\tau_{\pm}/dt=(2\pi\alpha_fc/\lambda_C)(m_ec^2/\hbar\omega_h)\chi{}T_{\pm}(\chi)$, are then solved until these optical depths are reached, at which point the emission event occurs \cite{Duclous_11}.  Here, $\alpha_f$ is the fine structure constant and $\lambda_C$ is the Compton wavelength; $\chi=(\hbar\omega_h/2m_ec^2)|\mathbf{E}_{\perp}+\hat{\mathbf{k}}\times{}c\mathbf{B}|$, where $\perp$ signifies the field component perpendicular to the unit vector in the photon's direction of motion $\hat{\mathbf{k}}$.  Photons are generated with a random energy weighted by the synchrotron function $F(\eta,\chi)$ including Klein-Nishina corrections \cite{Erber_66}.  $\chi$ controls pair production via the function $T_{\pm}(\chi)\approx0.16K^2_{1/3}(2/3\chi)/\chi$.  The generated pairs are treated on an equivalent footing to the original electrons in the PIC code, the photons are treated as massless, chargeless macroparticles which propagate ballistically.   The pairs are included when the PIC code calculates the charge and current densities on the computational grid, so contribute to the electromagnetic fields that are used to calculate the QED rates at the next timestep; ensuring a self-consistent simulation.  
 
\begin{figure}
\centering
\includegraphics[scale=0.55]{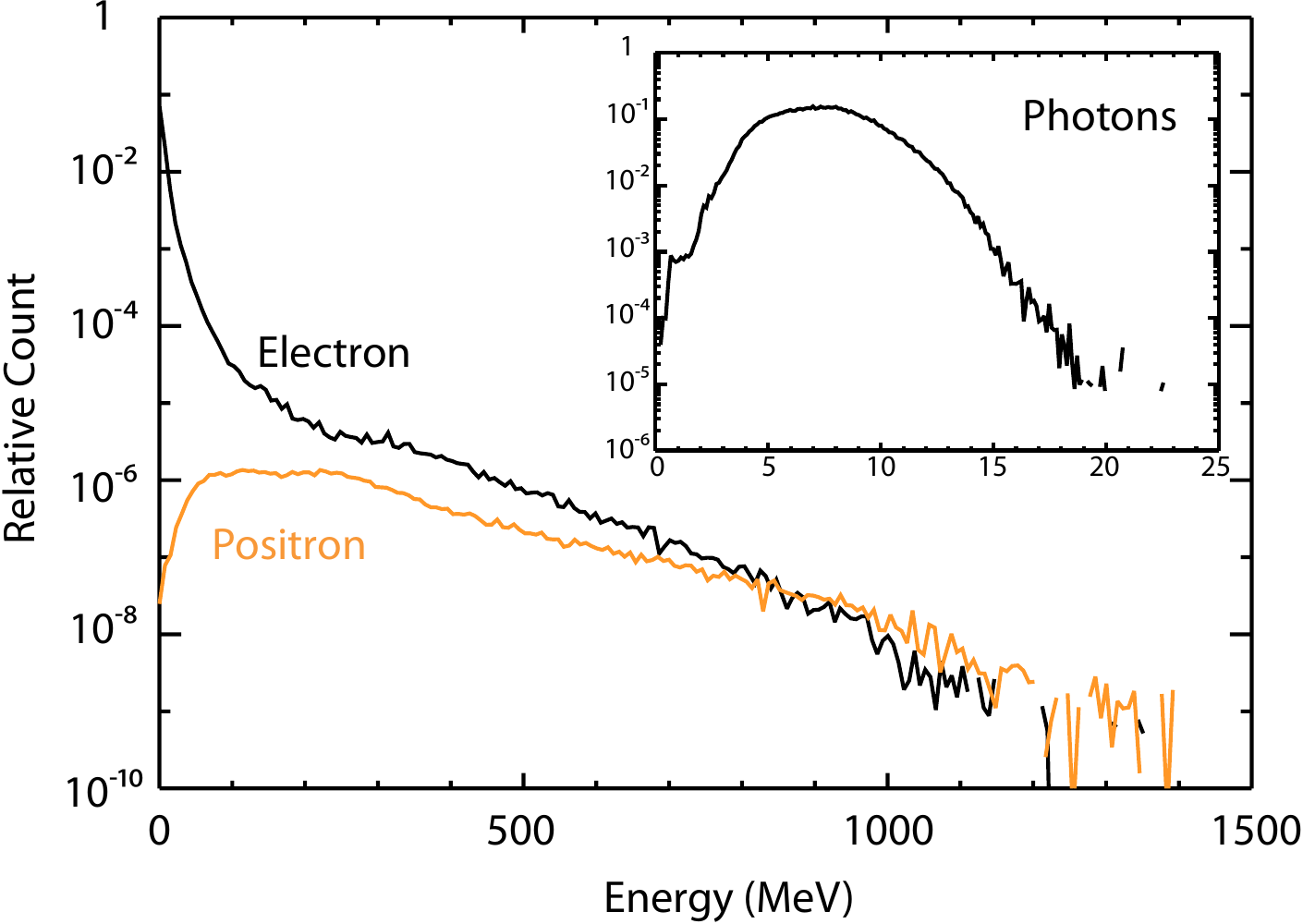}
\caption{\label{specs} Electron \& positron energy spectra in the interaction of a laser of intensity $4\times10^{23}$Wcm$^{-2}$ with solid aluminium (spatially and temporally integrated).  As usual the electron spectrum has a pronounced tail of `fast electrons'.  Corresponding gamma-ray spectrum (inset).}
\end{figure}

We have performed two-dimensional {\small{\small{EPOCH}}} simulations of a  10PW laser striking an aluminium foil including the QED processes.  The results are shown in Fig. \ref{solid_target_2D}.  The aluminium target is $1\mu$m thick, has a density of 2700kgm$^{-3}$ and is assumed to be fully ionised.  The target is represented by 1000 pseudo-electrons and 32 pseudo-ions per cell, with a spatial resolution of 10nm.  The laser has wavelength $\lambda_l=1\mu$m and is linearly p-polarised. The pulse has an energy of 377J and a duration of 30fs, with a square temporal profile.  It is focused to a spot of radius $1\mu{}$m with intensity $I=4\times10^{23}$Wcm$^{-2}$.  For this laser intensity the electron density of fully ionised aluminium $n_e$ is higher than the relativistically-corrected critical density $n_c=\gamma{}m_e\epsilon_0\omega_l^2/e^2$ ($\omega_l=2\pi{}c/\lambda_l$) and the plasma is overdense.  Therefore the laser beam is reflected and the light-pressure of the beam bores a hole into the target as shown in Fig. \ref{solid_target_2D}.  Also shown is prolific gamma-ray and positron production at the hole-boring front, where the laser is reflected.  The total number of pairs produced is $N_{\pm}=8\times10^9$ (each red dot is a macroparticle representing $2\times10^6$ positrons). 

Pairs are overall electrically neutral and so readily escape the target.  Thin sheets of pure electron-positron plasma form behind the target with a positron number density of $10^{26}$m$^{-3}$.  An $e^-e^+$ plasma is also trapped inside the hole-boring cavity with density $10^{25}$m$^{-3}$ over one cubic micron, forming a self-contained `micro-laboratory' potentially useful for the study of such a plasma.  For the 1$\mu$m thick target the laser just breaks through the target at the end of the 30fs laser pulse, releasing the trapped pairs for probing.  When the laser breaks through the situation reverts to that of a single electron in a single beam, pair production ceases and further laser energy is wasted.   The positron density is seven orders of magnitude higher than produced by the gold-target scheme described above and is high enough that collective effects could be studied with a CO$_2$ laser.  Fig. \ref{specs} shows that the average positron energy of 250MeV is much higher than the energy of photons from which they originate.  This suggests that the positrons are accelerated to high energy by the laser.  In this case we expect the average Lorentz factor of the positrons to be $\langle\gamma\rangle=a^{sol}+\Phi=2a^{sol}=2eE_{HB}^{sol}/m_ec\omega_l\approx$300MeV, which is consistent with the simulations.   $E_{HB}^{sol}$ is the value of the electric field inside the solid (as discussed below).  $\Phi$ is the sheath potential generated by fast electrons as they leave the target.  The sheath field acts to confine the fast electrons (the majority species compared to positrons) inside the target and so accelerates positrons \cite{Chen_10}, doing work approximately equal to the fast electron energy \cite{Ridgers_11}).  In practise lasers can have a long timescale pre-pulse of lower intensity than the main pulse.  Such a pre-pulse may heat the target and cause it to expand prior to the arrival of the main pulse, generating a pre-plasma.  Additional simulations, similar to that discussed above, show that a small ($e$-folding distance $=1\mu$m) pre-plasma does not dramatically reduce the number of gamma-ray photons and pairs generated.  In fact, for the parameters explored here a pre-plasma actually enhances gamma-ray production by 10\%.  A full exploration of this enhancement over all parameter space, as well as the role of pre-plasma in gamma-ray and pair production for a laser pulse at oblique incidence is beyond the scope of this paper.    

\begin{figure}
\centering
\includegraphics[scale=0.42]{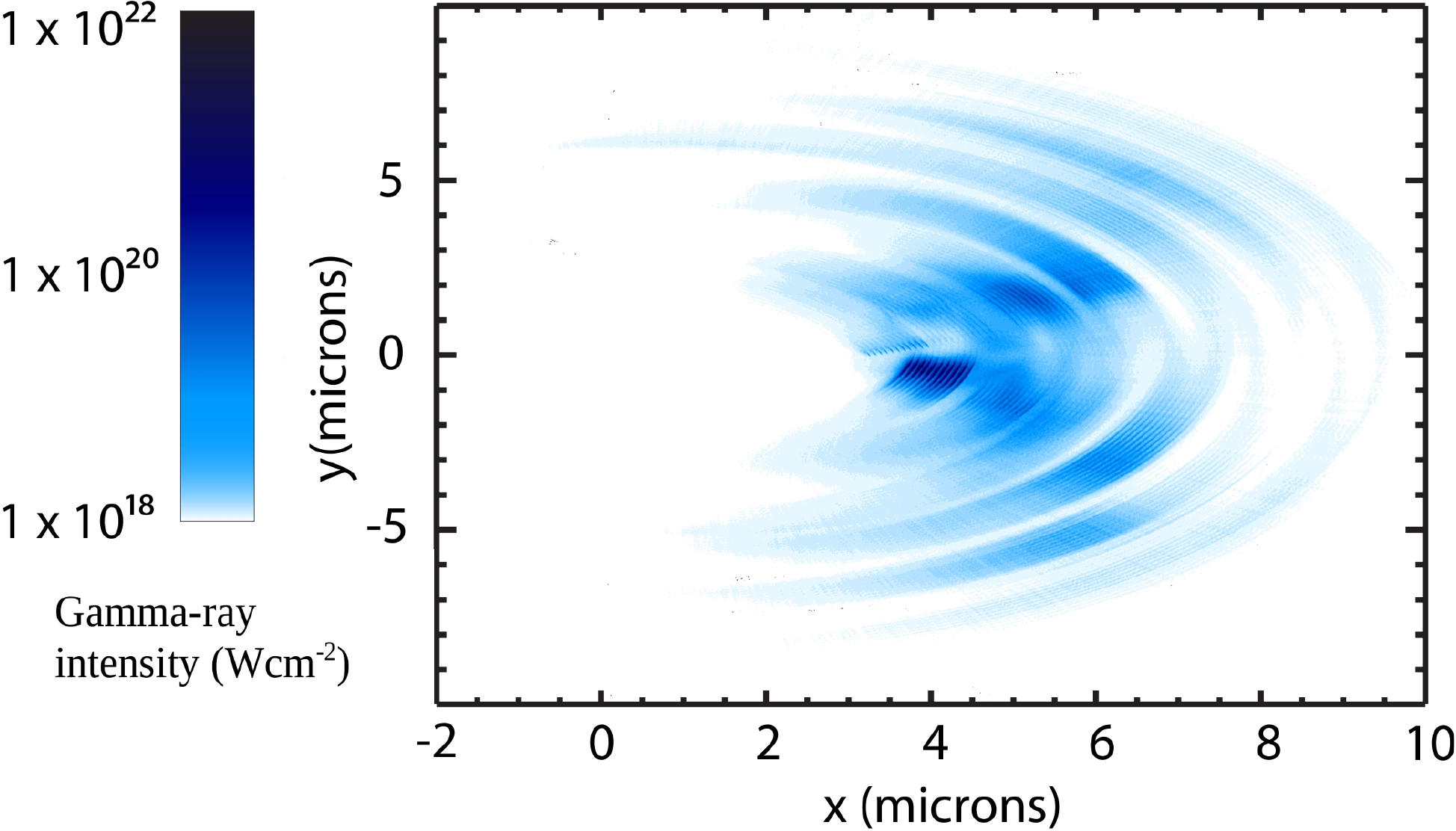}
\caption{\label{gamma_rays} Gamma-ray intensity averaged over one laser period for a laser intensity of $8\times10^{23}$Wcm$^{-2}$ 25fs after the end of the incident laser pulse (when all the photons leave the target).}
\end{figure}

Synchrotron gamma-ray photons are generated prolifically in the laser-solid interaction \cite{Zhidkov_02}.  At a laser intensity of $8\times10^{23}$Wcm$^{-2}$, a burst of gamma-rays of average intensity $8\times10^{21}$Wcm$^{-2}$ is produced at the rear of a 1$\mu$m thick Al target.  This is shown in Fig. \ref{gamma_rays}.  $10^{14}$ gamma-ray photons with an average energy of 16MeV are produced.  The conversion fraction of laser to gamma-ray energy is 0.35.  Note that synchrotron emission dominates over the more usual bremsstrahlung emission (not included in the simulation) during the laser pulse.  The synchrotron emission occurs during the pulse duration, whereas the bremsstrahlung cooling time (several picoseconds) is substantially longer.  In the simulation the synchrotron gamma-ray emission is contained within a cone half-angle of $\phi_{sim}=80^{\circ}$. which is consistent with the relativistically boosted angle $\phi_{boost}=\cos^{-1}(v_{HB}/c)$.  Here $v_{HB}$ is the speed of the hole-boring front (discussed further below).  The substantial energy loss to gamma-ray emission profoundly alters the energy budget of the laser-solid interaction and so the plasma physics processes.  The average electron energy is reduced from 41MeV to 21MeV \cite{Chen_11}.  The average ion energy is not strongly modified but the spectrum is substantially modified by the synchrotron emission; altering it from a single peak at 3.5GeV (the hole boring model, described below, predicts 3GeV) to two peaks at 1.5GeV \& 4.5GeV.  The reduced peak is due to the reduction in the reflected laser intensity, reducing the strength of the pistoning of the ion surface from $2I/c$ for perfect reflection to $I/c$ for perfect absorption.  The enhanced energy peak occurs as the laser breaks through, as show by Tamburini \emph{et al} \& Chen \emph{et al} \cite{Chen_11,Tamburini_10}. 

We have shown that gamma-ray emission alters the energy budget of laser-solid interactions and so the classical plasma physics. Conversely, the rates of the QED reactions are strongly modified by the plasma physics processes, closing the feedback loop which is the defining feature of QED-plasmas.  The modification of the QED rates can be estimated by employing the analytical model of Bell \& Kirk \cite{Bell_08}.  Here the controlling parameter $\eta$ is expressed implicitly in terms of $I_{24}$, the laser intensity in units of $10^{24}$Wcm$^{-2}$, and $\lambda_{\mu\mbox{\small{m}}}$, the laser wavelength in microns as $I_{24} = 2.75\eta^4+0.28\eta/\lambda_{\mu{}\mbox{{\small{m}}}}$.  In laser-solid interactions three plasma effects reduce $\eta$ and consequently the QED rates.  (1) \emph{Relativistic hole-boring} \cite{Robinson_09}: when $I<8\times10^{23}$Wcm$^{-2}$ the laser reflects from the overdense solid's hole-boring surface, which is moving at relativistic speed $v_{HB}$.  Where $v_{HB}/c=\sqrt{\Xi}/(1+\sqrt{\Xi})$ and $\Xi=I/\rho{}c^3$ is the dimensionless pistoning parameter.  The energy of the accelerated ions (of mass $m_i$) is $2\Xi{}m_ic^2/(1+2\sqrt{\Xi})$.  In the rest-frame of the hole boring surface, the intensity and wavelength are modified by the relativistic Doppler effect to $I_{24,HB}$ \& $\lambda_{\mu\mbox{\small{m}},HB}$  (2) \emph{The skin-effect}: the maximum value of the electric field in the evanescent wave inside the solid is reduced to $E^{sol}=2(n_c/n_{eHB})^{1/2}E_{HB}^{max}$.  $E_{HB}^{max}$ is the peak laser electric field and $n_{eHB}$ is the electron number density in the hole-boring frame.  This can be included by modifying $I_{24,HB}$ to $I_{24,HB}^{sol}=I_{24,HB}n_c/n_{eHB}$.  Maximum $I_{24,HB}^{sol}$ is achieved by reducing the target density such that it is just above the relativistic critical density at the incident laser intensity.  (3)  \emph{Self-induced transparency} \cite{Kaw_70}: for $I>8\times10^{23}$Wcm$^{-2}$ the solid target begins to become transparent, the situation approaches that of a laser interacting with a single electron and the rate of pair production is strongly reduced.

The equation for $\eta$ including plasma physics effects ($\eta_{HB}^{sol}$) is $I_{24,HB}^{sol}=2.75(\eta_{HB}^{sol})^4+0.28\eta_{HB}^{sol}/\lambda_{\mu{}\mbox{\small{m}},HB}$.  We can solve this numerically to obtain $\eta_{HB}^{sol}$ in a given laser-solid interaction and so estimate the number of gamma-ray photons produced and their energy.  To do this we use the rate equation for $d\tau_{\gamma}/dt$ given above with $\eta\rightarrow\eta_{HB}^{sol}$ and $F(\eta,\chi)\rightarrow{}f_{mono}(4\chi/3\eta^2=y)=(8\pi/9\sqrt{3})\delta(y-0.29)$.  The latter corresponds to assuming that the emitted photons are monochromatic with energy $\langle\hbar\omega_h\rangle=0.44\eta_{HB}^{sol}\langle\gamma\rangle{}m_ec^2$ \cite{Bell_08}.  The number of gamma-ray photons produced per electron per laser period is then given by $N_{\gamma}=6.42\alpha_f\langle\gamma\rangle$ \cite{Bell_08}.  $\langle\gamma\rangle$ is the average Lorentz factor of the electrons and can be estimated by $\langle\gamma\rangle\approx{}a^{sol}$.  For a laser of intensity $8\times10^{23}$Wcm$^{-2}$ focused onto a solid aluminium target $\eta_{HB}^{sol}\approx0.4$ and so $N_{\gamma}\approx4\times10^{13}$ and $\hbar\omega_h\approx25$MeV.  These are in reasonable agreement with the simulation results presented above.

An alternative configuration for pair production, recently investigated by Nerush \emph{et al} \cite{Nerush_11} \& Elkina \emph{et al} \cite{Elkina_11}, is the interaction of counter-propagating lasers in an underdense gas.  In this case plasma physics effects do not reduce the QED rates, but the plasma density is much lower.  In order to compare these configurations we performed one-dimensional {\small{\small{EPOCH}}} simulations of: (1) a laser of intensity $I$ striking a solid, semi-infinite (to avoid complicating break-through effects), Al target; (2) counter-propagating lasers of intensity $I/2$ in an underdense hydrogen gas-jet.  For $I<8\times10^{23}$Wcm$^{-2}$ more pairs are produced  by the solid target configuration (above this intensity the aluminium target becomes transparent).  The $10^5$ times denser plasma outweighs the $10^3-10^4$ times reduced rate of reaction for the solid.  This rate reduction can be estimated analytically.  The number of pairs produced per electron per laser period is $N_{\gamma}(1-e^{-\langle\tau\rangle})$ \cite{Bell_08}, where $\langle\tau\rangle$ is the photon optical depth for absorption over a distance $\lambda_l$.  Here $\langle\tau\rangle=12.8I_{24}e^{-4/3\langle\chi\rangle}$ and $\langle\chi\rangle=(\langle\hbar\omega_h\rangle/2m_ec^2)(E_{HB}^{sol}/E_s)$.  For a $I=8\times10^{23}$Wcm$^{-2}$ laser-aluminium interaction the reduction in $\eta$ leads to a reduction in $N_{\pm}$ by a factor of $10^4$, in good agreement with the simulations.  For intensities of the order of $10^{24}$Wcm$^{-2}$ (expected to be reached by 100PW class lasers) the gas-jet configuration produces more pairs.  In contrast to the solid a large fraction of the pairs generated go on to produce additional pairs, the reaction runs away and a cascade of antimatter production ensues.  This is in good agreement with the results of Nerush \emph{et al} \cite{Nerush_11}.  

In conclusion, we have shown that 10PW laser-solid interactions will generate dense electron-positron plasmas and ultra-intense bursts of gamma-rays, relevant to the laboratory study of pair production in high-energy astrophysical environments.  In contrast to the other laser-based positron production schemes mentioned, we have shown that for 10PW laser-solid interactions there is a strong feedback between QED processes and plasma physics, leading to the new regime of QED-plasma physics.  An understanding of future experiments in this regime will be impossible without a self-consistent model including the interplay between QED and classical plasma physics as discussed here.  

We acknowledge the support of the Centre for Scientific Computing, University of Warwick and thank Professor Peter Norreys \& Dr Brian Reville for useful discussions.   This work was funded by EPSRC grant numbers EP/GO55165/1 and EP/GO5495/1.


\begin{thebibliography}{99}
\bibitem{Goldreich_69} P. Goldreich, \& W.H. Julian, ApJ., \textbf{157}, 869 (1969)
\bibitem{Blandford_77} R.D. Blandford, \& R.L. Znajek, MNRAS, \textbf{179}, 433 (1977)
\bibitem{Ritus_85} V.I. Ritus, J. Russ. Laser Res., \textbf{6}, 5 (1985)
\bibitem{Timhokin_10} A.N. Timokhin, MNRAS, \textbf{408}, 2092 (2010)
\bibitem{Schwinger_51} J. Schwinger, Phys. Rev., \textbf{82}, 664 (1951)
\bibitem{Burke_97} D.L. Burke, \emph{et al}. Phys. Rev. Lett., \textbf{79}, 1626 (1997)
\bibitem{Sokolov_10} I.V. Sokolov, N.M. Naumova, J.A. Nees \& G.A. Mourou, Phys. Rev. Lett., \textbf{105}, 195005 (2010)
\bibitem{Hu_10} H. Hu, C. M\"{u}ller \& C.H. Keitel, Phys. Rev. Lett., \textbf{105}, 080401 (2010)
\bibitem{Chen_10} H. Chen, \emph{et al}. Phys. Rev. Lett., \textbf{105}, 015003 (2010)
\bibitem{Liang_98} E.P. Liang, S.C. Wilks \& M. Tabak, Phys. Rev. Lett, \textbf{81}, 4887 (1998)
\bibitem{Shen_01} B. Shen, \& J. Meyer-ter-Vehn, Phys. Rev. E, \textbf{65}, 016405 (2001) 
\bibitem{Bell_08} A.R. Bell, \& J.G. Kirk, Phys. Rev. Lett., \textbf{101}, 200403 (2008)
\bibitem{Fedotov_10} A.M. Fedotov, N.B. Narozhny, G. Mourou, \& G. Korn, Phys. Rev. Lett., \textbf{105}, 080402 (2010)
\bibitem{Bulanov_10} S.S. Bulanov, T.Z. Esirkepov, A.G.R. Thomas, J.K. Koga, \& S.V. Bulanov, Phys. Rev. Lett., \textbf{105}, 220407 (2010)
\bibitem{Nerush_11} E.N. Nerush, \emph{et al}, Phys. Rev. Lett., \textbf{106}, 035001 (2011)
\bibitem{Mourou_07} G.A. Mourou, C.L. Labaune, M. Dunne, N. Naumova \& V.T. Tikhonchuk, Plasma Phys. Control. Fusion, \textbf{49}, B667 (2007)  
\bibitem{Kirk_09} J.G. Kirk, A.R. Bell \& I. Arka, Plasma Phys. Control. Fusion, \textbf{51}, 085008 (2009)
\bibitem{Erber_66} T. Erber, Rev. Mod. Phys., \textbf{38}, 626 (1966)
\bibitem{Mackenroth_11} F. Mackenroth \& A. Di Piazza, Phys. Rev. A, \textbf{83}, 032106 (2011)
\bibitem{DiPiazza_10} A. DiPiazza, K.Z. Hatsagortsyan \& C.H. Keitel, Phys. Rev. Lett., \textbf{105}, 220403 (2010)
\bibitem{Brady_11} C.S. Brady \& T.A. Arber, Plasma Phys. Control. Fusion, \textbf{53}, 015001 (2011)
\bibitem{Shen_72} C.S. SHen \& D. White, Phys. Rev. Lett., \textbf{28}, 7 (1972)
\bibitem{Dirac_38} P.A.M. Dirac, Proc. R. Soc. A, \textbf{167}, 148 (1938)
\bibitem{Landau_87} L.D. Landau \& E.M. Lifshitz, The Course of Theoretical Physics Vol. 2 (Butterworth-Heinemann, Oxford, 1987) 
\bibitem{Sokolov_09} I.V. Sokolov, N.M. Naumova, J.A. Nees, G.A. Mourou \& V.P. Yanovsky, Phys. Plas., \textbf{16}, 093115 (2009)
\bibitem{Harvey_11} C. Harvey, T. Heinzl, N. Iji \& K. Langfeld, Phys. Rev. D, \textbf{83}, 076013 (2011)
\bibitem{Duclous_11} R. Duclous, J.G. Kirk \& A.R. Bell, Plasma Phys. Control. Fusion, \textbf{53}, 015009 (2011)
\bibitem{Ridgers_11} C.P. Ridgers, M. Sherlock, R.G. Evans, A.P.L. Robinson \& R.J. Kingham, Phys. Rev. E, \textbf{83}, 036404 (2011)
\bibitem{Zhidkov_02} A. Zhidkov, J. Koga, A. Sasaki \& M. Uesaka, Phys. Rev. Lett., \textbf{88}, 185002 (2002)
\bibitem{Chen_11} M. Chen, A. Pukhov, T. Yu \& Z. Sheng, Plasma Phys. Control. Fusion, \textbf{53}, 014004 (2011)
\bibitem{Tamburini_10} M. Tamburini, F. Pegoraro, A. Di Piazza, C.H. Keitel \& A. Macchi, NJP, \textbf{12}, 123005 (2010)
\bibitem{Robinson_09} A.P.L. Robinson \emph{et al}, Plasma Phys. Control. Fusion, \textbf{51}, 024004 (2009)
\bibitem{Kaw_70} P. Kaw \& J. Dawson, Phys. Fluids, \textbf{13}, 472 (1970)
\bibitem{Elkina_11} N.V. Elkina \emph{et al}, Phys. Rev. ST. AB., \textbf{14}, 054401 (2011)
\end{thebibliography}
\end{document}